\documentclass[aps,prb,twocolumn]{revtex4}
\usepackage{makeidx}
\usepackage{amsmath,amssymb,amsfonts,amsthm}
\usepackage{graphicx,bm}

\begin{document}

\title{Effects of coupling between octahedral tilting and polar modes on the
phase diagram of PbZr$_{1-x}$Ti$_{x}$O$_{3}$}
\date{}
\author{F. Cordero,$^{1}$ F. Trequattrini,$^{2}$ F. Craciun$^{1}$ and C.
Galassi$^{3}$}
\affiliation{$^1$ CNR-ISC, Istituto dei Sistemi Complessi, Area della Ricerca di Roma -
Tor Vergata,\\
Via del Fosso del Cavaliere 100, I-00133 Roma, Italy}
\affiliation{$^{2}$ Dipartimento di Fisica, Universit\`{a} di Roma \textquotedblleft La
Sapienza\textquotedblright , P.le A. Moro 2, I-00185 Roma, Italy}
\affiliation{$^{3}$ CNR-ISTEC, Istituto di Scienza e Tecnologia dei Materiali Ceramici,
Via Granarolo 64, I-48018 Faenza, Italy}

\begin{abstract}
The results are presented of anelastic and dielectric spectroscopy
measurements on large grain ceramic PbZr$_{1-x}$Ti$_{x}$O$_{3}$ (PZT) with
compositions near the two morphotropic phase boundaries (MPBs) that the
ferroelectric (FE) rhombohedral phase has with the Zr-rich antiferroelectric
and Ti-rich FE tetragonal phases. These results are discussed together with
similar data from previous series of samples, and reveal new features of the
phase diagram of PZT, mainly connected with octahedral tilting and its
coupling with the polar modes. Additional evidence is provided of what we
interpret as the onset of the tilt instability, when is initially frustrated
by lattice disorder, and the long range order is achieved at lower
temperature. Its temperature $T_{\mathrm{IT}}\left( x\right) $\ prosecutes
the long range tilt instability line $T_{\mathrm{T}}\left( x\right) $\ up to
$T_{\mathrm{C}} $, when $T_{\mathrm{T}}$ drops. It is proposed that the
difficulty of seeing the expected $\frac{1}{2}\left\langle 111\right\rangle $
modulations in diffraction experiments is due to the large correlation
volume associated with that type of tilt fluctuations combined with strong
lattice disorder.

It is shown that the lines of the tilt instabilities tend to be attracted
and merge with those of polar instabilities. Not only $T_{\mathrm{IT}}$\
bends toward $T_{\mathrm{C}}$ and then merges with it, but in our series of
samples the temperature $T_{\mathrm{MPB}}$\ of the dielectric and anelastic
maxima at the rhombohedral/tetragonal MPB does not cross $T_{\mathrm{T}}$,
but deviates remaining parallel or possibly merging with $T_{\mathrm{T}}$.
These features, together with a similar one in $\left( \text{Na}_{1/2}\text{%
Bi}_{1/2}\right) _{1-x}$Ba$_{x}$TiO$_{3}$, are discussed in terms of
cooperative coupling between tilt and FE instabilities, which may trigger a
common phase transition. An analogy is found with recent simulations of the
tilt and FE transitions in multiferroic BiFeO$_{3}$ [Kornev and Bellaiche,
Phys. Rev. B \textbf{79}, 100105 (2009)].

An abrupt change is found in the shape of the anelastic anomaly at $T_{%
\mathrm{T}}$ when $x$ passes from $0.465$ to 0.48, possibly indicative of a
rhombohedral/monoclinic boundary.
\end{abstract}

\maketitle


\section{Introduction}

The phase diagram of the most widely used ferroelectric perovskite PbZr$%
_{1-x}$Ti$_{x}$O$_{3}$ (PZT) still has unclear features (for the phase
diagram, see Fig. \ref{fig PD}). It has been known since the fifties\cite%
{Saw53,Jaf62,JCJ71} and the major recent discovery was the existence of a
monoclinic (M) phase\cite{NCS99} in a narrow region at the morphotropic
phase boundary (MPB) that separates the ferroelectric (FE) Zr-rich
rhombohedral (R) region from the Ti-rich tetragonal (T) one. In the M phase
the polarization may in principle continuously rotate between the directions
in the T and R domains, so providing an additional justification for the
well known and exploited maximum of the electromechanical coupling at the
MPB. The existence of domains of M phase is actually still debated, the
alternative being nanotwinned R and/or T domains that over a mesoscopic
scale appear as M.\cite{Kha10,SSK07b} Since experimental evidences for both
types of structures exist, the possibility should be considered that genuine
M domains and nanotwinning coexist at the MPB, being both manifestations of
a free energy that becomes almost isotropic with respect to the polarization.%
\cite{145} The part of the MPB line below room temperature has been
investigated only after the discovery of the M phase, and is reported to go
almost straight to 0~K at $x\simeq 0.52$.\cite{NCS00,SLA02}

Recent studies are also revealing new features of how the TiO$_{6}$ and ZrO$%
_{6}$ octahedra tilt at low temperature. The instability of the octahedral
network toward tilting is a common phenomenon in perovskites ABO$_{3}$,
usually well accounted for by the mismatch between the network of B-O bonds
with that of A-O bonds which are softer and with larger thermal expansion.%
\cite{Bro92,BDM97} In these cases, lowering temperature or increasing the
average B size sets the stiff B-O network in compression, which is relieved
by octahedral tilting.\cite{Meg46,Woo97b,RCS94,WKR05} In the case of PZT, Zr
has a radius 19\% smaller than Ti and one expects the tilt instability to
occur below a $T_{\mathrm{T}}\left( x\right) $ line that encloses the low-$T$
and low-$x$ corner of the $x-T$ phase diagram. Indeed, the Zr-rich
antiferroelectric compositions are tilted ($a^{-}a^{-}c^{0}$ in Glazer's
notation,\cite{Gla72} meaning rotations of the same angle about two
pseudocubic axes in anti-phase along each of them and no rotation around the
third axis), below a $T_{\mathrm{AF}}\left( x\right) $ line that goes
steeply toward 0~K at $x\sim 0.05$. Also at higher Ti compositions tilting
is observed ($a^{-}a^{-}a^{-}$ compatible with the rhombohedral $R3c$
structure) below a $T_{\mathrm{T}}$ line that presents a maximum at $x\sim
0.16$ and whose prosecution to low temperature was not followed beyond $%
x=0.4 $ until recently. The prediction from first principles calculations
that octahedral tilting occurs also in the M and T phase\cite{KBJ06} has
been confirmed by anelastic and dielectric,\cite{127} structural,\cite{HSK10}
Raman\cite{DFT11} and infra-red\cite{BNP11} experiments. The presence of a
low-temperature monoclinic $Cc$\ phase \cite{RSR05} with tilt pattern $%
a^{-}a^{-}c^{-}$\ intermediate between tilted R\ and T\ has been excluded by
a recent neutron diffraction experiment on single crystals,\cite{PLX10}
where below $T_{\mathrm{T}}$\ coexistence was found of tilted $R3c$\ and
untilted $Cm$\ phases. Yet, evidence for the $Cc$ phase has been
subsequently reported on PZT where 6\% Pb was substituted with smaller Sr,
in order to enhance tilting.\cite{SKM11}

Here we report the results of anelastic and dielectric experiments at
additional compositions with respect to our previous investigations, which
reveal new features of the phase diagram of PZT, mainly related to
octahedral tilting and its coupling with the polar degrees of freedom.

\section{Experimental}

Large grain (average sizes $15-30$~$\mu $m) ceramic samples of PbZr$_{1-x}$Ti%
$_{x}$O$_{3}$, with nominal compositions $x=0.05$, 0.062, 0.08, 0.12, 0.40,
0.487, 0.494 have been prepared with the mixed-oxide method in the same
manner as previous series of samples.\cite{127,145} The starting oxide
powders were calcined at 800~$^{\circ }$C for 4 hours, pressed into bars and
sintered at 1250~$^{\circ }$C for 2~h, packed with PbZrO$_{3}$\ + 5wt\%
excess ZrO$_{2}$\ to prevent PbO loss during sintering. The powder X-ray
diffraction did not reveal any trace of impurity phases and the densities
were about 95\% of the theoretical ones. The sintered blocks were cut into
thin bars $4~$cm long and $0.6$~mm thick for the anelastic and dielectric
experiments and discs with a diameter of 13~mm and a thickness of 0.7~mm
were also sintered only for the dielectric measurements. The faces were made
conducting with Ag paste.

The dielectric susceptibility $\chi \left( \omega ,T\right) =\chi ^{\prime
}-i\chi ^{\prime \prime }$ was measured with a HP 4194 A impedance bridge
with a four wire probe and an excitation of 0.5 V/mm, between 0.1 and
100~kHz. The heating and cooling runs were made at $0.5-1.5$~K/min between
100 and 800~K in a modified Linkam HFS600E-PB4 stage and up to 540~K in a
Delta climatic chamber.

The dynamic Young's modulus $E\left( \omega ,T\right) =E^{\prime
}+iE^{\prime \prime }\ $was measured between 100 and 770~K in vacuum by
electrostatically exciting the flexural modes of the bars suspended on thin
thermocouple wires.\cite{135} The reciprocal of the Young's modulus, the
compliance $s=s^{\prime }-is^{\prime \prime }=$ $1/E$, is the mechanical
analogue of the dielectric susceptibility. During a same run the first three
odd flexural vibrations could be excited, whose frequencies are in the
ratios $1:5.4:13.2$. The angular frequency of the fundamental resonance is%
\cite{NB72} $\omega \propto \sqrt{E^{\prime }}$, and the temperature
variation of the real part of the compliance is given by $s\left( T\right)
/s_{0}\simeq $ $\omega _{0}^{2}/\omega ^{2}\left( T\right) $, where $\omega
_{0}$ is chosen so that $s_{0}$ represents the compliance in the
paraelectric phase. The imaginary parts of the susceptibilities contribute
to the losses, which are presented as $Q^{-1}=s^{\prime \prime }/s^{\prime }$
for the mechanical case and $\tan \delta =\chi ^{\prime \prime }/\chi
^{\prime }$ for the dielectric one.

\section{Results}

For clarity, we will consider separately the anelastic and dielectric
spectra with compositions in the range $0.05<x<0.2$, and those in the MPB
region. We will present the new data together with those already published
in Ref. \onlinecite{127} ($x=0.455$, 0.465, 0.48, and 0.53) and Ref. %
\onlinecite{145} ($x=0.1$, 0.14, 0.17, 0.42, 0.45, 0.452).

\subsection{Octahedral tilting below $T_{\mathrm{T}}$ and $T_{\mathrm{IT}}$:
$0.062<x<0.2$}

\begin{figure}[htb]
\includegraphics[width=8.5 cm]{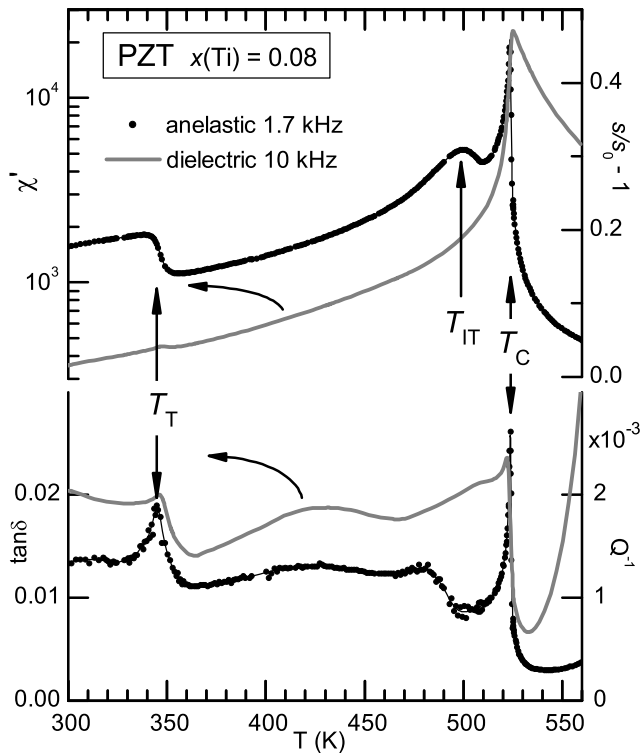}
\caption{Dielectric (left ordinates) and anelastic (right ordinates)\
spectra (higher panel real susceptibilities, lower panel losses) of PbZr$%
_{0.92}$Ti$_{0.08}$O$_{3}$ measured during heating.}
\label{fig Ti8}
\end{figure}

Figure \ref{fig Ti8} presents the dielectric and anelastic spectra measured
during heating of PbZr$_{0.92}$Ti$_{0.08}$O$_{3}$, a composition where also
the new transition at $T_{\mathrm{IT}}$ is clearly visible in the elastic
compliance $s^{\prime }$. The comparison between the two types of
susceptibilities puts in evidence their complementarity in studying
combinations of polar and non polar modes. The dielectric susceptibility $%
\chi ^{\prime }$ is of course dominated by the FE transition at $T_{\mathrm{C%
}}$ (note the logarithmic scale), it has a very attenuated step below the
well known tilt transition at $T_{\mathrm{T}}$, and practically nothing
visible at $T_{\mathrm{IT}}$, due to both the broader shape of the anomaly
and the proximity to the Curie-Weiss peak. The dielectric losses provide an
indirect but more clear mark of the non polar transition at $T_{\mathrm{T}}$%
, presumably through a change in the mobility and/or amplitude of charge and
polar relaxations which are affected by octahedral tilting.

The effect of cooling through $T_{\mathrm{T}}$ on the dielectric
susceptibility is more convincingly shown to be a positive step in Fig. \ref%
{fig dielTT}, and this fact will be discussed as a sign of cooperative
coupling between tilt and polar modes.

\begin{figure}[htb]
\includegraphics[width=8.5 cm]{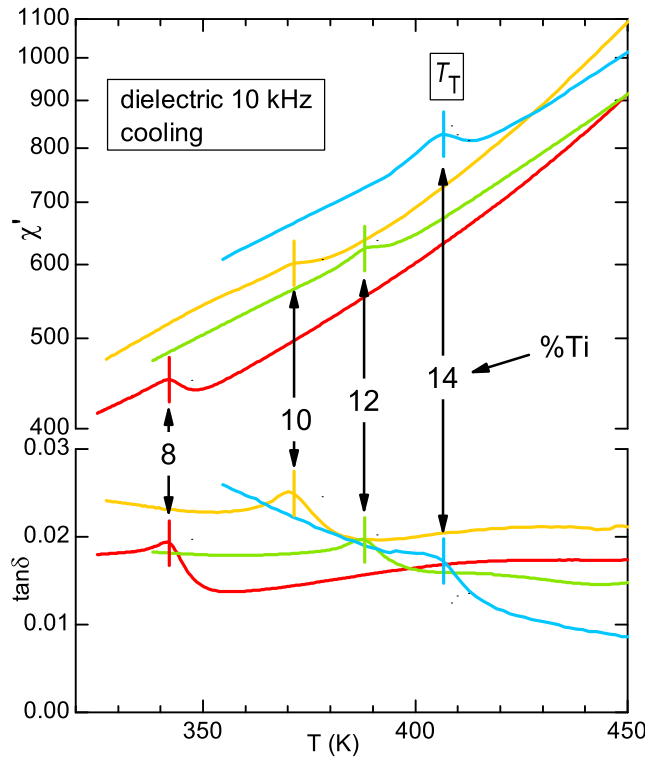}
\caption{(Color online) Dielectric susceptibility and loss of PbZr$_{1-x}$Ti$%
_{x}$O$_{3}$ measured during cooling through the tilt transition.}
\label{fig dielTT}
\end{figure}

The elastic compliance $s^{\prime }$, on the other hand, is only indirectly
affected by the FE transition, since strain is not an order parameter of the
transition and is linearly coupled to the square of the polarization. The
Landau theory of phase transitions\cite{Reh73,CS98} predicts a step in $%
s^{\prime }$ for this type of coupling, which is indeed observed at higher
Ti compositions,\cite{127} but has a strong peaked component in Zr-rich PZT.
We do not have an obvious explanation for this peaked response, which is
frequency independent and intrinsic, but mechanisms involving dynamical
fluctuations of the order parameter coupled with strain are possible.\cite%
{CS98} The advantage of a reduced anelastic response to the FE instabilities
is that the other transitions are not as masked as in the dielectric case,
so that not only is the tilt transition at $T_{\mathrm{T}}$ clearly visible
as a step in $s^{\prime }$ and peak in $Q^{-1}$, but also the new transition
can be detected at $T_{\mathrm{IT}}$ even very close to $T_{\mathrm{C}}$. As
already discussed,\cite{145} this transition has all the features of the
transition at $T_{\mathrm{T}}$ but the associated anomaly is attenuated and
broadened, so providing further support to an explanation in terms of a
disordered precursor to the final long range tilt ordering below $T_{\mathrm{%
T}}$.

The broad peaks and steps in both the dielectric and anelastic losses below $%
T_{\mathrm{C}}$ have scarce reproducibility, which indicates their extrinsic
origin, namely the motion of domain walls and charged defects, whose state
depends on the thermal history. Instead, all the features indicated by
arrows are completely independent of the measuring frequency, temperature
rate and thermal history, and therefore are recognizable as intrinsic
effects due to the FE and tilt transitions. Hysteresis between heating and
cooling is observed due to the first order character of the transitions and
to the presence of domain walls relaxations. Examples of the differences
between the features that are intrinsic and stable and those that present
dispersion in frequency or are less reproducible have been reported
previously\cite{127,145} and are omitted here.

\begin{figure}[htb]
\includegraphics[width=8.5 cm]{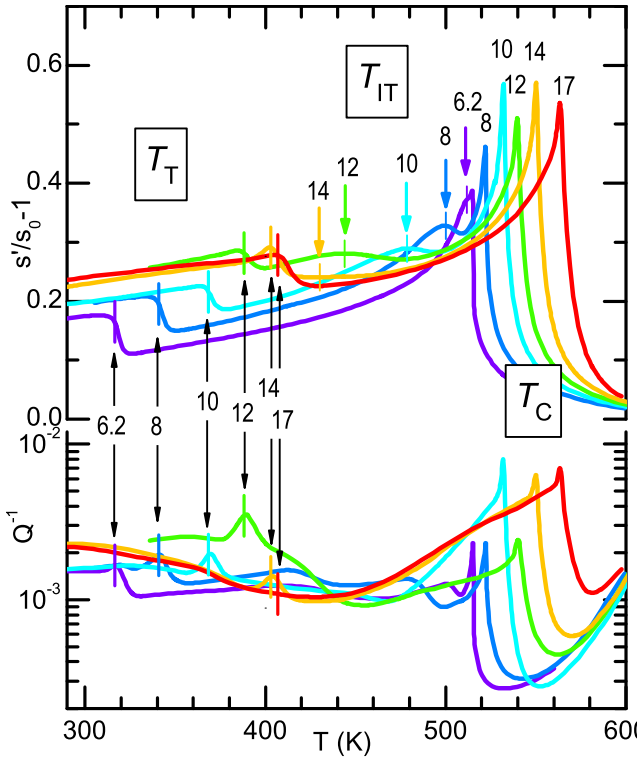}
\caption{(Color online) Elastic compliance $s^{\prime }$ and energy loss
coefficient $Q^{-1}$ measured at $\sim 1.7$~kHz on PZT at the compositions
6.2, 8, 10, 12, 14, 17\% Ti, as indicated by the numbers at the phase
transitions. The curves of 10, 14 and 17\% Ti are from Ref. \onlinecite{145}%
. }
\label{fig anel6-17}
\end{figure}


Figure \ref{fig anel6-17} presents the anelastic spectra of PZT with $%
0.062<x<0.17$, including compositions already present in Ref. %
\onlinecite{145}. All the curves are similar to the $x=0.08$ case of Fig. %
\ref{fig Ti8}, with the three type of transitions at $T_{\mathrm{C}}$, $T_{%
\mathrm{IT}}$ and $T_{\mathrm{T}}$ clearly visible in separate temperature
ranges. Both the $s^{\prime }$ and $Q^{-1}$ curves have sharp peaks at the
FE transitions, so that the $T_{\mathrm{C}}$'s are simply labeled with the
compositions in \%Ti. The other transition temperatures are indicated by
vertical bars centered on the curves and arrows labeled with the respective
compositions. The features of $Q^{-1}$ in the $T_{\mathrm{IT}}$ temperature
range are not labeled because are due to the extrinsic contributions
mentioned above.

\begin{figure}[htb]
\includegraphics[width=8.5 cm]{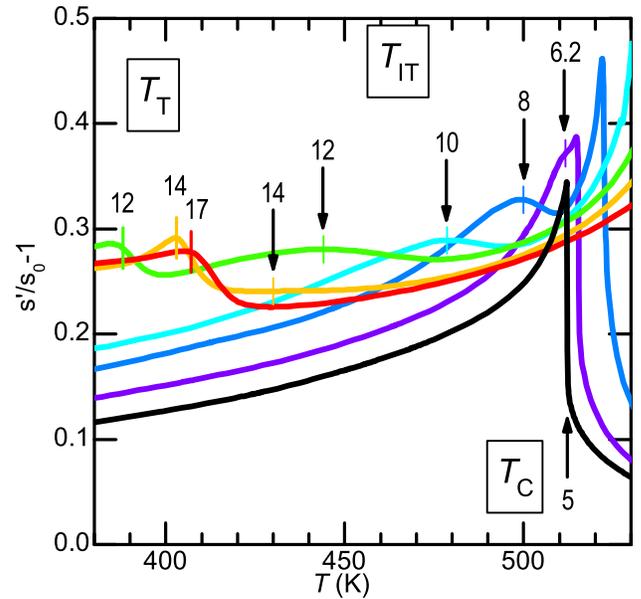}
\caption{(Color online) Detail of the anomalies of the elastic compliance at
$T_{\mathrm{IT}}$, measured at $\sim 1.7$~kHz during cooling on samples with
$0.05\leq x\leq 0.17$. The numbers indicate the compositions in \%Ti. The
curves of 10, 14 and 17\% Ti are from Ref. \onlinecite{145}.}
\label{fig s5-17}
\end{figure}

The temperatures of the tilt transition are identified with the upper edges
of the steps in the real parts, which generally coincide with a spike or
sharp kink in the losses. The rounded step and lack of reproducible anomaly
in the losses increase the error on $T_{\mathrm{IT}}$, which however remains
small enough to not change the features of the phase diagram discussed
later. Due to the importance of the behavior of $T_{\mathrm{IT}}\left(
x\right) $ in the Discussion, a detail of this anomaly in the $s^{\prime
}\left( T\right) $ curves, including 5\% Ti, is shown in Fig \ref{fig s5-17}%
. An anomaly corresponding to $T_{\mathrm{IT}}$ might be present slightly
above $T_{\mathrm{T}}$ for $x=0.17$, but lacking a clear sign of it, it is
assumed to coincide with $T_{\mathrm{T}}$. At low $x$, the curve of $x=0.05$
does not present any clear shoulder below $T_{\mathrm{C}}$, and it is
assumed $T_{\mathrm{IT}}\equiv T_{\mathrm{C}}$.

The transition temperatures measured on both heating and cooling are
reported in the phase diagram of Fig. \ref{fig PD}, where the $T_{\mathrm{IT}%
}$ line departs from $T_{\mathrm{T}}$ at $x\simeq 0.17$, has a kink centered
at $x=0.11$ and finally joins the $T_{\mathrm{C}}$ line at $0.05<$ $x<0.062$%
. The new feature that will be the main focus of the present work is the
kink and the merging with $T_{\mathrm{C}}$ at $x>0.05$. It is also
noticeable that the anomaly at $T_{\mathrm{IT}}$ becomes more intense and
sharper on approaching $T_{\mathrm{C}}$.

\subsection{Compositions near the MPB}

As discussed in the previous investigations,\cite{127,145} the MPB in PZT is
signaled by a maximum in the dielectric and above all elastic
susceptibilities. Again, it is stressed that such maxima are almost
independent of frequency and temperature rate, and therefore are intrinsic
effects due to the evolution of the order parameter at the MPB and its
coupling with strain. Also the losses are rather high in the region of the
MPB, but no feature is found that is directly ascribable to the phase
transition; rather, their dependence on frequency and thermal history show
that they are due to the abundant twin walls and other domain boundaries,
whose density and mobility depend on many factors and is maximal around the
MPB. Instead, the losses contain clear cusps or steps at the tilt
transitions,\cite{127,145} so allowing $T_{\mathrm{T}}$ to be determined
also in the proximity with the MPB, where the real part is dominated by the
peak at $T_{\mathrm{MPB}}$. We therefore discuss separately the real parts
of $\chi $ and $s$, containing information on the polar transition at the
MPB, and the losses, containing information on the tilt transitions.

\subsection{Maxima of the susceptibilities at the MPB}

\begin{figure}[htb]
\includegraphics[width=8.5 cm]{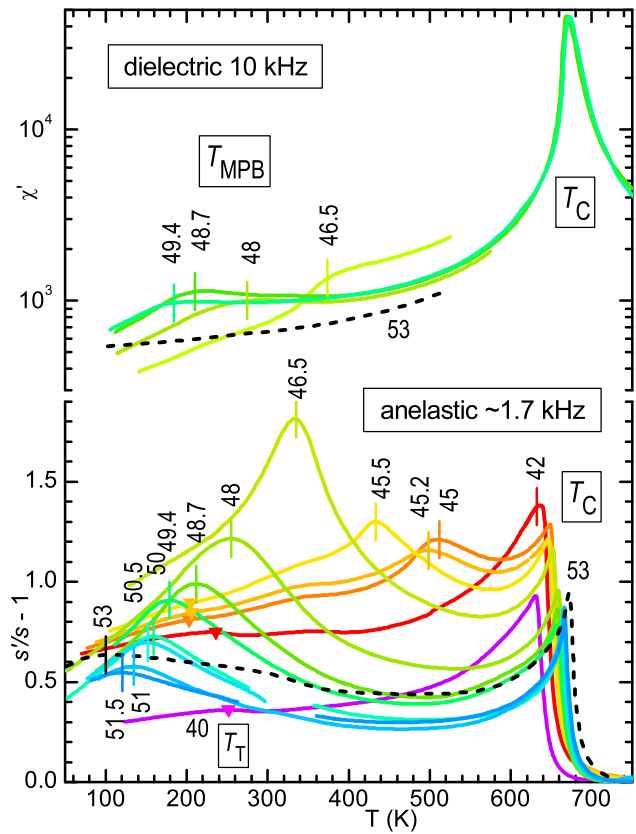}
\caption{(Color online) Dielectric susceptibility and elastic compliance
measured during cooling on PZT at the compositions indicated besides the
curves in \% Ti. The $T_{\mathrm{MPB}}$'s are indicated with vertical bars
and $T_{\mathrm{T}}$ by triangles (only for $40\leq $ $x\leq $ $45.2$).
Present work: 40, 48.7 and 49.4\% Ti; the other curves are from Refs.
\onlinecite{145} and \onlinecite{127}.}
\label{fig 40to53}
\end{figure}

Figure \ref{fig 40to53} is an overview of $\chi ^{\prime }$ and $s^{\prime }$
curves measured during cooling at all the compositions $x\geq 0.40$ we
tested so far.

We call $T_{\mathrm{MPB}}$ the temperatures of the maxima in $\chi ^{\prime
} $ and $s^{\prime }$, marked with vertical bars on the curves. These
temperatures do not coincide exactly with each other, because $\chi ^{\prime
}$ and $s^{\prime }$ are two different response functions of polarization
and strain respectively, but, once plotted in the phase diagram, they
present an excellent correlation with the MPB determined by diffraction, at
least in the middle of the MPB line (see Fig. \ref{fig PD}). The dielectric
maxima at $T_{\mathrm{MPB}}$ are much smaller and broader than the
Curie-Weiss peak at $T_{\mathrm{C}}$ (note the logarithmic scale), whereas
the anelastic maxima at $T_{\mathrm{MPB}}$ have comparable or even larger
intensities than the step at $T_{\mathrm{C}}$ (part of the peaked component
at $T_{\mathrm{C}}$ has a frequency dispersion denoting relaxation of walls%
\cite{127,145}).

At $x=0.40$ there is no peak attributable to the MPB, but only a minor step
below $T_{\mathrm{T}}$, which is indicated with triangles up to $x=0.452$;
beyond that composition, the step at $T_{\mathrm{T}}$ either disappears or
is masked by the MPB peak. The other shallow anomaly centered at $\sim 360$%
~K in the curves up to $x\leq 0.465$ is the counterpart of the domain wall
relaxation appearing in the losses mentioned above and will be ignored. For $%
x\geq 0.45$ the peak at the MPB shifts to lower temperature and develops its
maximum amplitude at 0.465, which has been argued to correspond to the point
of the phase diagram where the anisotropy of the free energy is minimum.\cite%
{145} The presence of a peak in $s^{\prime }$ at the MPB has also been
argued to be evidence that the phase transition occurring at the MPB
consists mainly in the rotation of the polarization, from the $\left[ 001%
\right] $ direction of the T phase toward the $\left[ 111\right] $ direction
of the R\ phase. In fact, in that case the transverse (perpendicular to the
original $\left[ 001\right] $ direction) component of $\mathbf{P}$ acts as
order parameter and is almost linearly coupled to a shear strain, inducing a
peaked response also in the elastic susceptibility.\cite{127,145} This would
be an evidence that a monoclinic phase, and not only nanotwinned R and T
phases, exists below the MPB. Yet, the smooth shape of the maximum is
compatible with an inhomogeneous M phase coexisting and possibly promoted by
nanotwinning.\cite{145} In fact, indications continue to accumulate of
intrinsic phase heterogeneity near the MPB compositions also on single
crystals.\cite{BBF12}

Beyond $x>0.465$, the peak at $T_{\mathrm{MPB}}$ gradually decreases its
amplitude and temperature, and, thanks to the great number of closely spaced
compositions, is clearly recognizable as the signature of the MPB up to $%
x=0.515$. The next composition, $x=0.53$ (dashed curves), still has a
shallow maximum at a temperature that prosecutes the $T_{\mathrm{MPB}}\left(
x\right) $ line, but its nature appears different. In fact, the dielectric $%
\chi ^{\prime }$ at the same composition lacks any sign of a maximum, and
the overall $s^{\prime }$ curve does not any more prosecute the trend of the
preceding curves. For this reason, the temperature of this maximum at $%
x=0.53 $ is reported in the phase diagram as $T_{\mathrm{MPB}}$ but
accompanied by a question mark. A $T_{\mathrm{MPB}}$ is extracted also from
the curve at $x=0.42$, even though a separate maximum is not present. It is
however the only composition where $s^{\prime }$ has no sharp feature at $T_{%
\mathrm{C}}$, and we assume that this is due to a rounded peak at $T_{%
\mathrm{MPB}}$ very close to $T_{\mathrm{C}}$.

\subsection{Tilt transition near the MPB}

\begin{figure}[htb]
\includegraphics[width=8.5 cm]{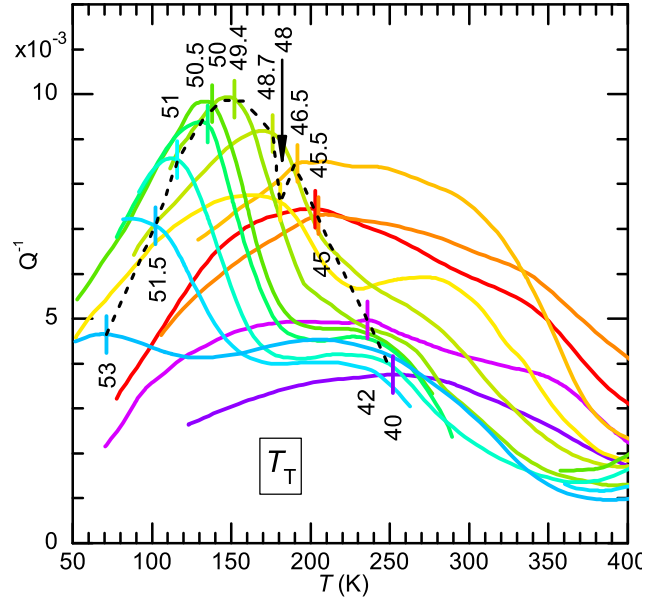}
\caption{(Color online) Elastic energy loss coefficient of PZT at the
compositions indicated by the numbers (in \%Ti), measured during cooling at $%
\sim 1.7$~kHz. The anomalies at $T_{\mathrm{T}}$ are indicated by vertical
bars and joined with a dashed line. Present work: 40, 48.7 and 49.4\% Ti;
the other curves are from Refs. \onlinecite{145} and \onlinecite{127}.}
\label{fig TT20-53}
\end{figure}

The best signatures of the tilt transition below $T_{\mathrm{T}}$ are found
in the anelastic losses. Figure \ref{fig TT20-53} shows the $Q^{-1}\left(
T\right) $ curves at all the compositions $x\geq 0.40$ we tested so far
(only 45.2\%Ti is omitted in order to not overcrowd the figure).

The $T_{\mathrm{T}}$'s up to $x=0.455$ are the same as deduced from the step
in the real part and indicated by triangles in Fig. \ref{fig 40to53}. Up to $%
x=0.465$ $T_{\mathrm{T}}$ is identified with the temperature of a spike in $%
Q^{-1}\left( T\right) $, which gradually becomes a cusp and starting from $%
x=0.48$ becomes a large step. As in the previous figures, the $T_{\mathrm{T}%
} $'s are marked by vertical bars centered on the curves and joined by a
dashed line, in order to better follow the evolution of the anomaly. The
transition between the spike/cusp and the step anomaly is unexpectedly
sudden, since it occurs within $0.465<x<0.48$. Such a discontinuity appears
also in the dashed line joining the anomalies, and is marked by an arrow. We
emphasize again that the losses generally have a limited reproducibility,
because depend on the status of domain walls and defects; therefore, the
regularity of the dashed curve joining the tilt anomalies of so many
different samples is remarkable and testifies the good and uniform quality
of the samples.

\section{Discussion}

\begin{figure}[htb]
\includegraphics[width=8.5 cm]{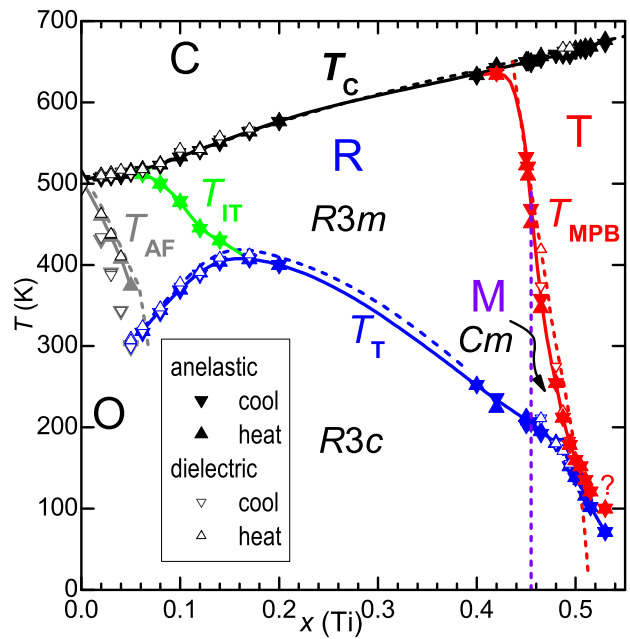}
\caption{(Color online) Phase diagram of PZT based on our anelastic and
dielectric spectra. The solid lines join the anelastic data measured during
heating; the dashed lines are those from Jaffe and Noheda. The question mark
reminds that the shallow maximum of $s^{\prime }$ with $x=0.53$ at that
temperature probably does not signal the MPB crossing.}
\label{fig PD}
\end{figure}

We refer to the phase diagram of PZT in Fig. \ref{fig PD}. Below $T_{\mathrm{%
C}}$ and with decreasing Ti content, one finds the following phases:\cite%
{YZT09,ZYG11b,WKR05} ferroelectric (FE) tetragonal (T) $P4mm$ with
polarization $\mathbf{P}$ along $\left[ 001\right] $, monoclinic (M) $Cm$
with $\mathbf{P}$ rotated toward $\left\langle 111\right\rangle $,
rhombohedral (R) $R3m$ with $\mathbf{P}\parallel \left\langle
111\right\rangle $ and antiferroelectric (AFE) orthorhombic (O) $Pbam$ with
staggered cations shifts along $\left\langle 110\right\rangle $ and $%
a^{-}a^{-}c^{0}$ tilt pattern. Below $T_{\mathrm{T}}$ octahedral tilting
occurs in all phases.

In Fig. \ref{fig PD}, the solid lines join the transition temperatures
deduced from our anelastic spectra measured during heating (filled triangles
pointing upward), which are generally very close to the points deduced from
the dielectric curves (empty triangles). The temperatures measured during
cooling are also shown as triangles pointing downward. The figure contains
all the data presented here and in Refs. \onlinecite{127,145} and, for
completeness, also points obtained at compositions $x\leq 0.05$, that will
be discussed in a future paper.

The dashed lines are from the most widely published version of Jaffe, Cook
and Jaffe\cite{JCJ71} with modifications of Noheda \textit{et al.}\cite%
{NCS00} around the MPB. In a different version\cite{FES86} the $T_{\mathrm{%
MPB}}$ line below $x=0.45$ does not prosecute straight up to join $T_{%
\mathrm{C}}$ almost perpendicularly, but rapidly decreases its slope and
joins $T_{\mathrm{C}}$ at $x\sim 0.3$. At present we have indication of a
much smaller deviation of $T_{\mathrm{MPB}}$ from the datum at $x=0.42$
(Fig. \ref{fig 40to53}), but already at $x=0.4$ there is no trace of a
double transition.

\subsection{The octahedral tilt instability\label{octa-inst}}

The tendency of the BO$_{6}$ octahedra in A$^{2+}$B$^{4+}$O$_{3}$
perovskites to tilt has been widely studied and can be rationalized in terms
of mismatch between a softer sublattice of longer A-O bonds that compresses
a stiffer sublattice of shorter B-O bonds, until the incompressible
octahedra tilt in order to accommodate a reduction of the cell and
cuboctahedral volume $V_{A}$ without reducing their volume $V_{B}$. In the
majority of cases the A-O bonds are softer than the B-O bonds, because they
are longer and the cation A shares its valence with 12 nearest neighbors O
atoms while B with only 6 of them.\cite{Bro92}

This effect is quantitatively expressed in terms of the Goldschmidt
tolerance factor
\begin{equation*}
t=\frac{\overline{\text{A-O}}}{\sqrt{2}\overline{\text{B-O}}}\simeq \frac{r_{%
\mathrm{A}}+r_{\mathrm{O}}}{\sqrt{2}\left( r_{\mathrm{B}}+r_{\mathrm{O}%
}\right) }
\end{equation*}%
which is 1 for the cubic untilted case. In order to predict a tendency to
form a tilted phase, $t$ is written in terms of the ideal ionic radii $r_{%
\mathrm{X}}$ of the appropriate valence and coordination, which are
tabulated.\cite{Sha76} If $t<1$ then the equilibrium B-O bond lengths are
too long to fit in a cubic frame of A-O bonds, and tilting occurs below some
threshold value; for example, many Sr/Ba based perovskites are tilted when $%
t<0.985$ at room temperature.\cite{RCS94} Alternatively, the polyhedral
volume ratio $V_{A}/V_{B}$ is defined,\cite{Tho89} which is 5 for the
untilted case and becomes $<5$ upon tilting. Of all the normal modes of a
cubic perovskite, those that induce a decrease of $V_{A}/V_{B}$ are the
combinations of rigid rotations of the octahedra about the cubic axes, while
all other normal modes involve distortions of the polyhedra with little or
no change in their volumes.\cite{WA11} Rotations with all the octahedra in
phase along the axis are labeled $M_{3}$, because the staggered shifts of
the O atoms create a modulation with vector $\frac{1}{2}\left\langle
110\right\rangle $, the M point of the Brillouin zone, while rotations with
successive octahedra in antiphase along the axis are labeled $R_{4}$,
because the modulation is $\frac{1}{2}\left\langle 111\right\rangle $, the R
point of the Brillouin zone.\cite{HS98} These modes are also called
antiferrodistortive (AFD), because of the staggered modulation of the atomic
displacements, and do not cause the formation of electric dipoles. Most of
the low temperature structures of perovskites can be described in terms of
combinations of these rotations together with polar modes (shifts of the
cations against the O anions). In particular, the low-temperature $R3c$
rhombohedral phase of PZT, is obtained from the high temperature $R3m$
ferroelectric R phase, by applying anti-phase rotations of the same angle
about all three pseudocubic axes ($a^{-}a^{-}a^{-}$ in Glazer's notation\cite%
{Gla72}), while the O-AFE structure $Pbam$ of PbZrO$_{3}$ is a combination
of staggered and hence AFE\ displacements of the Pb$^{2+}$ and Zr$^{4+}$
ions along $\left[ 110\right] $, with anti-phase tilts along $\left[ 100%
\right] $ and $\left[ 010\right] $ ($a^{-}a^{-}c^{0}$).

In spite of the considerable amount of research on the phase diagram of PZT,
little use has been done of these concepts in order to interpret it. An
indication that the mechanism governing tilting in PZT is indeed the A-O/B-O
mismatch is the observation that the substitution of 6\% Pb$^{2+}$ with the
smaller Sr$^{2+}$ induces tilting in an originally untilted T phase of PZT,
while when codoping with Sr$^{2+}$ and Ba$^{2+}$, the latter larger than Pb$%
^{2+}$, the opposite effects of the two dopants on the average tolerance
factor cancel with each other and no tilting is found.\cite{ZRL02} When the
tilt boundary $T_{\mathrm{T}}$ has been discussed in terms of $t$, the
incongruence of the deep depression near the border with the O-AFE phase has
been pointed out,\cite{WKR05} and explained in terms of frustration of AFE
displacements of the Pb ions perpendicularly to the average FE direction $%
\left\langle 111\right\rangle $. Such displacements lack the long range
order of the AFE-O structure, and their frustration would be transmitted to
the octahedral tilting through the Pb-O bonds, so lowering the $T_{\mathrm{T}%
}$ border in proximity with the AFE-O\ phase.\cite{WKR05} Certainly the
sharp depression in the border to the long range ordering of tilts, the $T_{%
\mathrm{AF}}+T_{\mathrm{T}}$ line, appears to be caused by some frustration,
since it is a typical feature of the phase diagrams with competing states,%
\cite{BMM01b,Dag05b} and both the FE/AFE polar modes and different tilt
patterns may be in competition.

Let us first consider the AFE/FE competition. The disordered displacements
of Pb away from the average polarization have been proposed to occur over
the whole R region of the phase diagram and particularly near the other MPB
with the T phase, based on the large anisotropic displacement ellipsoids of
Pb in Rietveld refinements according to the R structure\cite{CGW98} or to
the coexistence of M and R structures.\cite{YZT09,PLX10} While in these
cases the Pb displacements off-axis with respect to the $\left\langle
111\right\rangle $ direction may be imagined as having a FE correlation
leading to a rotation of the polarization away from $\left\langle
111\right\rangle $, recently a soft mode corresponding to AFE Pb
displacements has been found near the MPB of PZT and in relaxor PMN and
PZN-PT, suggesting that the AFE-like instability is a common feature of
nanoscale domain structures of rhombohedral or pseudorhombohedral lead-based
perovskites.\cite{HOK11} Considering that such displacements occur over the
whole R region or particularly at the MPBs with the O and the T phases, if
they are so strongly coupled with the tilts, one would expect a depression
of $T_{\mathrm{T}}$ also near the MPB with the T phase, which is not
observed. There are some possible explanations for the different behavior of
$T_{\mathrm{T}}$ at the two MPBs. One is that the polar displacements away
from $\left\langle 111\right\rangle $ might become static and with larger
amplitude and AFE correlation only near the AFE border, so causing the
frustration between FE and AFE patterns, while at the MPB to the T-FE phase
the AFE correlations remain at the stage of an anomalous phonon softening.

On the other hand, it is possible that the main competition occurs between
the two different tilt patterns in the R and O phases, since the approaching
of the tilt instability to the FE one indicates that in the $x\rightarrow 0$
region of the phase diagram polar and tilt modes have comparable energetics.
It is not evident, however, how the competition between $a^{-}a^{-}c^{0}$
and $a^{-}a^{-}a^{-}$ tilt patterns would cause frustration, since they can
transform into each other just by switching on and off the anti-phase tilt
about the pseudocubic axis $\left[ 001\right] $. We believe that so strong a
depression of the $T_{\mathrm{AF}}+T_{\mathrm{T}}$ lines is possible because
partial tilting has already occurred at $T_{\mathrm{IT}}$.

\subsection{The intermediate tilt instability $T_{\mathrm{IT}}\left(
x\right) $ line}

Before discussing the nature of the instability at $T_{\mathrm{IT}}$, we
emphasize the reasons why a positive step of the compliance during cooling
like that at $T_{\mathrm{IT}}$ should indicate a phase transition and not
some kinetic effect related to domain walls or defects: \textit{i)} cooling
causes pinning or freezing of domain walls and therefore decreases the
susceptibility, while an increase is observed at $T_{\mathrm{IT}}$; \textit{%
ii)} in the absence of tilting, the only conceivable walls just below $T_{%
\mathrm{C}}$ would be between the R-FE domains; if some anomaly in their
behavior occurred around $T_{\mathrm{IT}}$, it would appear also, or mainly,
in the dielectric susceptibility; \textit{iii)} the shape of the anomaly is
independent of the temperature rate, history and frequency,\cite{145} and
therefore is an intrinsic lattice effect.

The $T_{\mathrm{IT}}$ line appears as the prosecution of the $T_{\mathrm{T}}$
one toward the lowest $x$, or equivalently lowest $t$, and highest
temperature, and hence it has been identified as the onset of precursor
tilting.\cite{145} Presumably, between $T_{\mathrm{IT}}$ and $T_{\mathrm{T}}$
tilting would be disordered due to the enhanced disorder in the O-Pb-O
network near the AFE border, as suggested above.\cite{WKR05} The drastic
depression of $T_{\mathrm{T}}$ would then be due to the fact that most of
the mismatch between the Pb-O and (Zr/Ti)-O\ sublattices is relieved at $T_{%
\mathrm{IT}}$, and the final transition to a tilted phase with long range
order requires the buildup of sufficient elastic energy in the disordered
sublattice of tilted octahedra, that it is convenient to switch to the long
range ordered phase. This is not alternative to the above discussion on the
sharp minimum of $T_{\mathrm{AF}}+T_{\mathrm{T}}$ in terms of competing
phases. It is simply assumed that the frustration hinders tilting from
reaching long range order but not from occurring on the local scale.

This interpretation has not yet been corroborated by a structural study
where the onset of tilting below $T_{\mathrm{IT}}$ is actually observed and
the tilt pattern is established. Actually, superlattice spots in electron
diffraction have been interpreted in terms of in-phase $M_{3}$-type tilts
within the otherwise $R3m$ phase up to $x\sim 0.15$.\cite{Vie95,VLD96} That
region roughly corresponds to the region delimited by the $T_{\mathrm{IT}}$
border, though a border was not seen.\cite{Vie95,VLD96} These in-phase tilts
could not be confirmed by x-ray or neutron diffraction and considerable
debate ensued over the interpretation of the electron diffraction $\frac{1}{2%
}\left\langle 110\right\rangle $ spots in terms of AFE $\left\langle
110\right\rangle $ Pb displacements rather than $M_{3}$-type tilts, due to
the much weaker strength of the latter and the greater sensitivity of
electron diffraction to the damaged surface.\cite{RCW98,WKR05}

\begin{figure}[htb]
\includegraphics[width=8.5 cm]{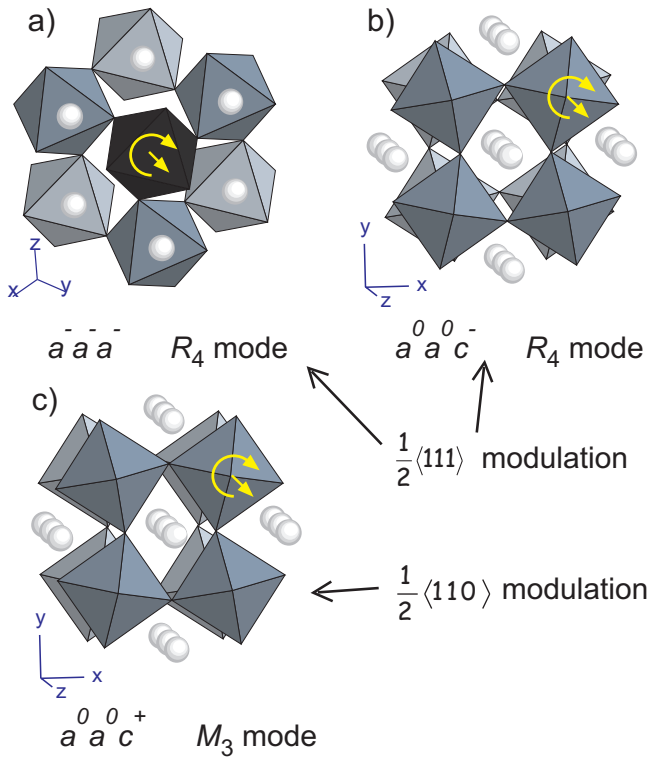}
\caption{Three tilt patterns of the BO$_{6}$ octahedra. The O ions at the
vertices of the octahedra and the B ions at their centers are not shown; the
A ions are white. The darker octahedra are in planes closer to the observer.
a) $a^{-}a^{-}a^{-}$ tilt pattern of the $R3c$ structure seen from a
direction close to the $\left\langle 111\right\rangle $ rotation axis; b)
out of phase and c) in phase rotation about the $\left[ 001\right] $
direction. In b) and c) the octahedra of different planes share only O atoms
that do not move under tilting..}
\label{fig tilts}
\end{figure}


The $M_{3}$ tilts are in the first place in the search for a mechanism
behind the phase transformation at $T_{\mathrm{IT}}$, because they would
cause weak superlattice peaks at positions coinciding with those from the
disordered AFE Pb displacements, so explaining why their onset below $T_{%
\mathrm{IT}}$ has not yet been noticed. In fact, the disorder in the Pb
sublattice, including AFE-like shifts, survives even in the cubic phase, as
demonstrated by the observation of electron diffraction $\frac{1}{2}%
\left\langle 110\right\rangle $ spots\cite{Vie95,VLD96} and the fact that
EXAFS probes the same local environment of Pb\cite{KTK05c} and Zr/Ti,\cite%
{RS95} both in the FE and in the cubic phases. It seems reasonable to assume
that the AFE cation displacements develop their short range order together
with the FE order below $T_{\mathrm{C}}$, so producing superlattice peaks of
the $\frac{1}{2}\left\langle 110\right\rangle $ type, that mask those from
subsequent octahedral tilting below $T_{\mathrm{IT}}$, if this is of $M_{3}$
type. The difficulty with the last assumption is that the final tilting
below $T_{\mathrm{T}}$ is of $R_{4}$ type, and the $M_{3}$ in-phase tilting
is an instability of different type rather than a precursor to it.

\subsection{$\frac{1}{2}\left\langle 110\right\rangle $ reflections from
hindered $\frac{1}{2}\left\langle 111\right\rangle $ modulation of the tilts?%
}

In order to reconcile the observation of $M$-type modulation below $T_{%
\mathrm{IT}}$ with an expected short range $R$-type modulation, we assume
that the propagation of tilt fluctuations is hindered by the lattice
disorder associated with Ti substituting Zr, with the consequent frustration
of the Pb displacements, and possibly other defects. Therefore, rather than
to consider fluctuations of infinitely extended normal modes, it is more
appropriate to consider the fluctuations of small clusters of octahedra,
whose size is limited by the local disorder. This makes the different types
of tilting represented in Fig. \ref{fig tilts} inequivalent, since they have
different correlation volumes. In a first approximation consider a network
of rigid octahedra, that, in order to comply with the mismatch with the Pb-O
network, can only tilt without distortions. This is the so-called rigid unit
model, whose implications on the anisotropy of the phonon dispersions has
been analyzed.\cite{SHD94} Here we focus on the effects that the anisotropy
of the correlation length might have on the diffraction patterns.

Consider first the rotation of a single octahedron about one of the cubic
axes ($\left[ 001\right] $ in Fig. \ref{fig tilts}b) or c)); this will cause
a rotation of all the other octahedra in the same plane perpendicular to the
axis in an AFD fashion, creating a $\frac{1}{2}\left\langle 110\right\rangle
$ modulation, with a correlation length $l_{\perp }$, perpendicular to the
rotation axis, which is an increasing function of the B-O bond strength. In
the pure rigid unit model $l_{\perp }$ is infinite and there is no
correlation at all with the other planes of octahedra, because they share
corners of the tilting octahedra only through immobile O atoms. In practice
there is an interaction with the adjacent planes through the less energetic
B-O-B bond bending and the extensions of the longer and weaker A-O\ bonds,
resulting in a finite correlation length $l_{\parallel }$ along the rotation
axis, however shorter than $l_{\perp }$. This means that the correlation
volume surrounding an octahedron tilting about a $\left\langle
100\right\rangle $ direction is a flat disc of diameter $2l_{\perp }$ and
thickness $2l_{\parallel }$ with $l_{\parallel }\ll l_{\perp }$, and this
will create superlattice reflections more intense and sharp at $\frac{1}{2}%
\left\langle 110\right\rangle $ than at $\frac{1}{2}\left\langle
111\right\rangle $. On the other hand, the rotation of an octahedron about a
$\left\langle 111\right\rangle $ axis (Fig. \ref{fig tilts}a)), as the soft $%
R_{4}$ mode in the $R3c$ structure, will affect all the surrounding
octahedra, which share shifted O atoms both above and below the $\left[ 111%
\right] $ plane containing the rotating octahedron. Therefore, for an
octahedron rotating about a $\left\langle 111\right\rangle $ axis the
correlation volume is a sphere with diameter $2l_{\perp }$, containing more
octahedra than in the previous case. We therefore postulate that below $T_{%
\mathrm{IT}}$ the magnitudes of the tilts start becoming so large to cause
their propagation through the correlation volume, as for any tilt
instability, but this will occur first for those clusters where a deviation
toward $\left\langle 100\right\rangle $ tilt, with smaller correlation
volume, is favored by the local cation disorder. In this first stage between
$T_{\mathrm{IT}}$ and $T_{\mathrm{T}}$, only $\frac{1}{2}\left\langle
110\right\rangle $ superlattice peaks would appear, easily masked by those
from AFE cation correlations. On further cooling below $T_{\mathrm{T}}$, the
increased instability of the $R_{4}$ mode and long range elastic
interactions trigger the long range tilt order of the $R3c$ phase. In other
words, in the absence of disorder or frustration $T_{\mathrm{T}}$ and $T_{%
\mathrm{IT}}$ would coincide and represent the temperature below which the
propagation of the $a^{-}a^{-}a^{-}$ tilt instability starts. In the
presence of disorder, local tilting about $\left\langle 100\right\rangle $
axes is favored because of its smaller correlation volume; the resulting
disordered tilting partially relieves the mismatch between Pb-O\ and
(Zr/Ti)-O bonds, and further cooling below $T_{\mathrm{T}}$ is necessary in
order to reach the long range $a^{-}a^{-}a^{-}$ ground state. This may
explain the deep depression of $T_{\mathrm{T}}$ below $x<0.15$.

\subsection{The kink in the tilt instability $T_{\mathrm{IT}}\left( x\right)
$ line and the effect of coupling of different modes on the phase diagram}

\begin{figure}[htb]
\includegraphics[width=8.5 cm]{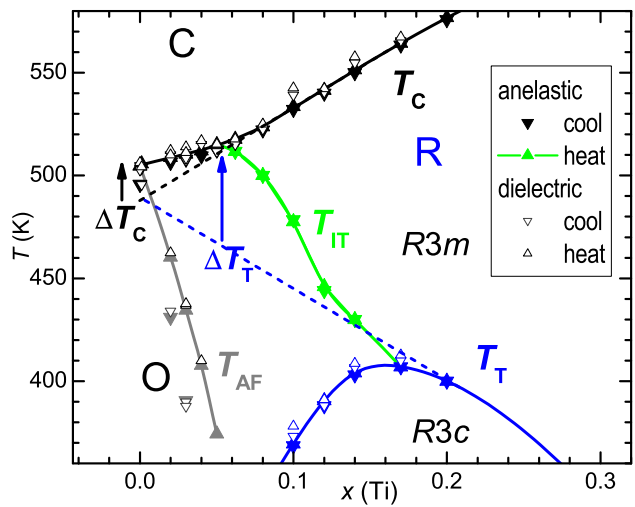}
\caption{(Color online) Zr-rich region of the phase diagram based on our
anelastic and dielectric spectra. The dashed lines are extrapolations from
the high-$x$ behavior of $T_{\mathrm{C}}\left( x\right) $ and $T_{\mathrm{T}%
}\left( x\right) $. The arrows $\Delta T_{\mathrm{C}}$ and $\Delta T_{%
\mathrm{T}}$ represent the deviation from the extrapolated behavior.}
\label{pdZrrich}
\end{figure}

If octahedral tilts and polar modes were independent of each other, the $T_{%
\mathrm{C}}$ and $T_{\mathrm{T}}+T_{\mathrm{IT}}$ lines might approach and
possibly cross each other in an independent manner. Instead, $T_{\mathrm{IT}%
} $ merges with $T_{\mathrm{C}}$ at $x=0.06$ with a noticeable kink around $%
x\sim 0.1$. The $T_{\mathrm{IT}}$ line seems "attracted" by $T_{\mathrm{C}}$%
, as if the tilt instability were favored by the ferroelectric one. On the
other hand, also $T_{\mathrm{C}}$ may feel some effect from the proximity
with $T_{\mathrm{IT}}$, since its slope slightly decreases after joining
with $T_{\mathrm{IT}}$ below $x\sim 0.05$. It appears that both $T_{\mathrm{%
IT}}$ and $T_{\mathrm{C}}$ increase with respect to the trend extrapolated
from higher $x$, when they are far from each other. This is shown in Fig. %
\ref{pdZrrich}, where the dashed lines are the extrapolations (with no
pretension of quantitative analysis) of $T_{\mathrm{C}}$ and $T_{\mathrm{T}}$%
. The vertical arrows are the maximum deviations of the actual $T_{\mathrm{C}%
}$ and $T_{\mathrm{IT}}$ lines from their extrapolations, $\Delta T_{\mathrm{%
C}}\sim 16$~K and $\Delta T_{\mathrm{T}}\sim 48$~K. A possible qualitative
explanation of this observations is in terms of cooperation between a
stronger FE instability and a weaker AFD tilt instability. The FE mode
leaves the lattice unstable also below $T_{\mathrm{C}}$, since its
restiffening is gradual, and also affects the modes coupled to it, in
particular favoring the condensation of modes cooperatively coupled to it at
a temperature higher than in the normal stiff lattice away from $T_{\mathrm{C%
}}$. The rotations of the octahedra are certainly coupled with the polar
modes, as demonstrated by the polarization\cite{WCG78,CNI97} and dielectric%
\cite{DXL95,145} anomalies at the tilt transitions, and the issue whether
this coupling is cooperative or competitive is discussed in Sect. \ref%
{comp/coop}. Assuming that the first eventuality is true, one has a
mechanism that enhances $T_{\mathrm{IT}}$ in the proximity of $T_{\mathrm{C}%
} $, but the coupling works also in the other sense: the proximity of the
tilt instability favors the condensation of the FE mode, enhancing $T_{%
\mathrm{C}} $. The energies involved in the FE instability are larger than
those involved in tilting, as indicated by the fact that $T_{\mathrm{C}}>T_{%
\mathrm{T}}$, and accordingly the perturbation of the FE mode on the tilt
mode is larger: $\Delta T_{\mathrm{T}}\sim 3\Delta T_{\mathrm{C}}$. A
phenomenological model\cite{Hol73} that may be applied to explain these
effects will be described in Sect. \ref{trigger}.

\subsection{Kinks in the $T_{\mathrm{MPB}}$ and $T_{\mathrm{T}}$ lines\label%
{kinks}}

\begin{figure}[htb]
\includegraphics[width=8.5 cm]{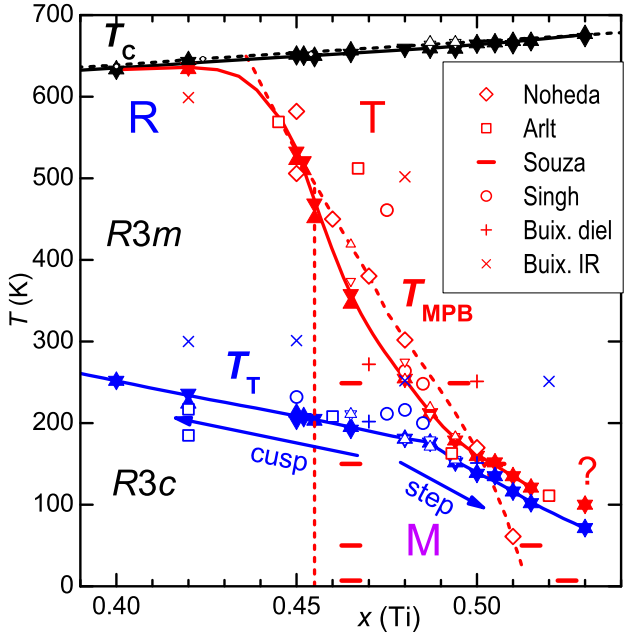}
\caption{(Color online) Enlargement of the MPB region of the phase diagram
of PZT. Triangles and solid lines from our measurements, as in Fig. \protect
\ref{fig PD}. Dashed lines and diamonds from Noheda and other open symbols
as indicated in the legend. The shape of the continuous $T_{\mathrm{MPB}%
}\left( x\right) $ line between $x=0.42$ and 0.45 is hypothetical.}
\label{pdMPB}
\end{figure}

Other new features of the PZT phase diagram that derive from the present
data are the approaches of the $T_{\mathrm{MPB}}\left( x\right) $ line with $%
T_{\mathrm{T}}$ and $T_{\mathrm{C}}$, and a distinct kink in $T_{\mathrm{T}}$
when encounters $T_{\mathrm{MPB}}$. This is better seen in the detail of the
MPB region in Fig. \ref{pdMPB}, where, besides the same data of Fig. \ref%
{fig PD}, other points of $T_{\mathrm{MPB}}$ and $T_{\mathrm{T}}$ are
reported from the literature. The data are from diffraction\cite{NCS00} ($%
\Diamond $) piezoelectric coefficient\cite{Arl90} $d_{11}$ ($\square $), $%
1/s_{11}$ measured with piezoelectric resonance\cite{SMR08} (o), Raman\cite%
{SLA02} (---), dielectric (+) and infrared ($\times $) \cite{BNP11}
spectroscopies.

Let us first consider how $T_{\mathrm{T}}\left( x\right) $ enters the MPB
region. The points from the literature, obtained from different techniques
and samples, are rather sparse, but those from our anelastic and dielectric
spectra have little dispersion, and show a clear change of slope of $T_{%
\mathrm{T}}\left( x\right) $ when it approaches $T_{\mathrm{MPB}}$ at $%
0.487< $ $x$ $<0.494$. This narrow composition range is close to but not the
same over which the anomaly in $Q^{-1}$ changes between cusped and steplike
(Fig. \ref{fig TT20-53}). In fact that change, marked by arrows in Fig. \ref%
{pdMPB}, occurs at $x\lesssim 0.48$, and therefore the two changes may
depend on different mechanisms. We will discuss the transition in the shape
of the $Q^{-1}$ anomaly in Sect. \ref{R/M}, and now we focus on the kink in
the $T_{\mathrm{T}}$ line, which we think is closely connected with the
proximity to the MPB.

Also our $T_{\mathrm{MPB}}$ points draw a curve with little dispersion,
compared to the body of data in the literature, but in this case a
difference emerges in the low temperature region: even though with only
three points below 100~K, the data from diffraction\cite{NCS00} and Raman%
\cite{SLA02} define an almost straight MPB border that ends at $T=0$ at $%
x=0.520\pm 0.005$. Instead, our data define a curved line that never crosses
$T_{\mathrm{T}}$. Our closely spaced points in the phase diagram and the
regular evolution of the spectra from which they are obtained (Figs. \ref%
{fig 40to53} and \ref{fig TT20-53}) suggest that the effect is real and
characteristic of good quality ceramic samples. The last point with the
question mark, obtained from the dashed curve in Fig. \ref{fig 40to53},
probably does not correspond to $T_{\mathrm{MPB}}$, but the difference
remains at $x=0.515$ between our curve at 120~K, and two points at $50-60$~K
from diffraction and Raman scattering. These discrepancies may depend on
differences in the samples rather than on the experimental technique. In
fact, the existence of the intermediate monoclinic phase and its nature are
not yet unanimously accepted, and it is also proposed that, besides
nanoscale twinning, defect structures like planes of O vacancies may have a
role in defining the microstructure of PZT and act as nuclei for
intermediate phases.\cite{RSR09} Hence, there is a range of microstructures
that may well reflect in the position of the MPB, but, again, the
consistency and regularity of the data encourage to consider the features
presented here as intrinsic of the PZT phase diagram and not vagaries from
uncontrolled defects.

It results that also $T_{\mathrm{T}}$ and $T_{\mathrm{MPB}}$ almost coincide
over an extended composition range, with $T_{\mathrm{MPB}}$ seemingly pushed
up by $T_{\mathrm{T}}$. For $x>0.49$, $T_{\mathrm{MPB}}$ deduced from the
maximum in $s^{\prime }$ and $T_{\mathrm{T}}$ deduced from the step in $%
Q^{-1}$ run parallel and close to each other and it is difficult to assess
whether they still represent two distinct transitions or instead they are
the manifestations of a same combined polar and tilt transition.

\subsection{Merging of tilt and polar instabilities also in NBT-BT\label%
{merging}}

\begin{figure}[htb]
\includegraphics[width=8.5 cm]{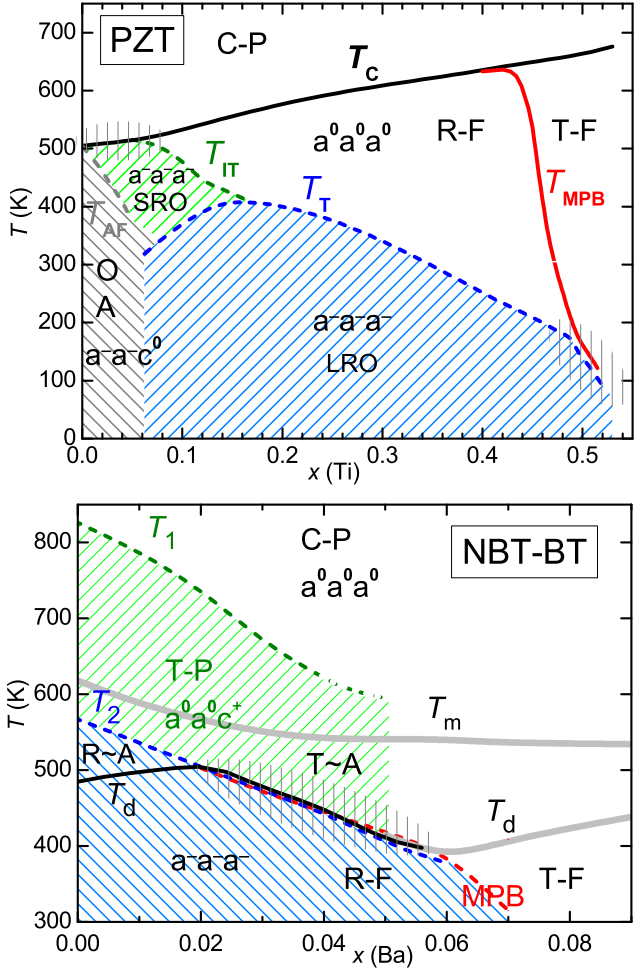}
\caption{(Color online) Phase diagrams of PZT and NBT-BT,\protect\cite{139}
where the broken lines are tilt instability borders, while the solid lines
are polar instability borders. Vertical hatching evidences the regions where
the two types of lines merge. Differently slanted patterns represent
different tilt patterns. P, F and A stand for paraelectric, ferroelectric
and antiferroelectric; SRO and LRO stand for short/long range order.}
\label{figPDs}
\end{figure}

Another example in which tilt instability lines merge with polar instability
lines is (Na$_{1/2}$Bi$_{1/2}$)$_{1-x}$Ba$_{x}$TiO$_{3}$ (NBT-BT).\cite{139}
This system has much stronger chemical disorder than PZT and a more
complicated and less defined phase diagram, especially beyond the MPB
composition $x\left( \text{Ba}\right) >0.06$, where the correlation lengths
are so short to render the material almost a relaxor. In Fig. \ref{figPDs}
the phase diagram of NBT-BT is presented together with that of PZT. The
broken lines represent the borders between regions with different tilts,
while the solid lines are polar instabilities ($T_{\mathrm{AF}}$ in PZT is
both tilt and polar). The regions were two types of instabilities merge are
vertically hatched, while differently slanted hatches represent different
tilt patterns. In NBT the tolerance factor is small due to the smallness of
the mean A ion size of Na$^{+}$ and Bi$^{3+}$ combined, and is increased by
substituting with Ba. Tilting occurs in two stages, first $a^{0}a^{0}c^{+}$
(T phase)\ below $T_{1}$ and then $a^{-}a^{-}a^{-}$ (R) below $T_{2}$, and
both $T_{1}$ and $T_{2}+T_{\mathrm{MPB}}$ have negative $dT/dx$, so
enclosing the low temperature/low tolerance factor region of the phase
diagram, as discussed in Sect. \ref{octa-inst} (the $T_{1}$ line actually
disappears into a highly disordered relaxor-like region). At variance with
PZT, the polar instabilities occur at temperatures lower than those of
tilting, and in two stages: first an almost AFE or ferrielectric region
below a temperature $T_{m}$ signaled by a maximum in the dielectric
susceptibility, and then FE below the so-called depolarization temperature $%
T_{\mathrm{d}}$. The intermediate $\sim $AFE structure is due to shifts of
the Ti and Bi cation along $\left[ 001\right] $ in opposite directions, so
to make $P$ almost null, and therefore is not the result of coupling with
the AFD $a^{0}a^{0}b^{+}$ tilting.\cite{JT02} The FE phase, however, appears
below a $T_{\mathrm{d}}<T_{2}$ at $x=0$, which rises and merges with the
decreasing $T_{2}\left( x\right) $ at $x=$ 0.02\cite{139} or $0.03$.\cite%
{HWN07}. Beyond 6\% Ba, the $T_{\mathrm{d}}$ and $T_{2}\equiv T_{\mathrm{MPB}%
}$, the latter ill defined, separate again. As a result, the border $T_{%
\mathrm{d}}$ to the FE phase, instead of simply crossing $T_{2}$, merges
with it in the range $0.02<x<0.06$, following a wavy path. This would not be
the only instance in the NBT-BT phase diagram where the tilt modes trigger a
mixed tilt-polar transition, since it has been recognized that in pure NBT
both the transitions at $T_{1}$\ and $T_{2}$\ are of such a type.\cite{PKF04}

Similarly to the cases of $T_{\mathrm{IT}}/T_{\mathrm{C}}$ and $T_{\mathrm{C}%
}/T_{\mathrm{MPB}}$ in PZT, the coincidence of $T_{\mathrm{d}}$ and $T_{2}$
in the composition range $0.02<x\left( \text{Ba}\right) <0.06$ may be
interpreted as a manifestation of cooperative coupling between tilt and FE
instabilities. Interestingly, like Pb$^{2+}$ also Bi$^{3+}$ has a lone pair
electronic configuration, with the tendency to reduce its coordination
number and form short bonds with covalent character\cite{CLW96,JT02} which
couple tilt and polar modes.\cite{GCW99}

\subsection{Merging of tilt and polar instabilities seen as trigger type
transitions\label{trigger}}

A possible mechanism for the merging of two transitions with order
parameters (OPs) of different symmetries had been proposed by Holakowsk\'{y}%
\cite{Hol73} and carried on by Ishibashi;\cite{Ish94b} This is also the case
of octahedral tilting, whose OP is the rotation angle $\omega $, and polar
or antipolar modes with OP $P$ (all one-dimensional for simplicity).
Holakowski\cite{Hol73} treated the case of a FE transition triggered by
another OP, and we will adapt his arguments to a tilt transition triggered
by the FE instability, referring to $T_{\mathrm{IT}}$. In this case the
minimal Landau expansion of the free energy is
\begin{equation}
F=\frac{a}{2}P^{2}+\frac{b}{4}P^{4}+\frac{a_{\omega }}{2}\omega ^{2}+\frac{%
b_{\omega }}{2}\omega ^{4}+\frac{c_{\omega }}{6}\omega ^{6}+F_{c}  \label{F}
\end{equation}%
where $a=\alpha \left( T-T_{\mathrm{C}}\right) $ and $a_{\omega }=\alpha
_{\omega }\left( T-T_{\omega }\right) $ represent the soft modes ideally
vanishing at $T_{\mathrm{C}}$ and $T_{\omega }<$ $T_{\mathrm{C}}$. In the
argument, the temperature dependence of $a_{\omega }$ is irrelevant, and
Holakowski sets it constant; in our case it should be $\lesssim T_{\mathrm{IT%
}}$). The coupling part $F_{c}$ contains mixed terms $P^{m}\omega ^{n}$.
Since $\omega $ and $P$ have different symmetries, not all the mixed terms
are invariant under the allowed symmetry operations in the cubic phase,
under which $F$ must be invariant; the lowest order term allowed by symmetry
is\cite{HFJ89}
\begin{equation}
F_{c}=-\frac{\gamma }{2}P^{2}\omega ^{2}~.  \label{Fc}
\end{equation}%
Such a term is always possible but generally overlooked when mixed terms of
lower order prevail. If $\gamma >0$ this term lowers the free energy when
both $P\neq 0$ and $\omega \neq 0$, describing a cooperative
polarization-tilting coupling. After solving the equilibrium condition $0=$ $%
\partial F/\partial P$, the equilibrium value of $\omega $ is expressed in
terms of equilibrium $P$, so that $F$ is written only in terms of $P$,\ and $%
F_{c}$ renormalizes $b$ as $b^{\prime }=b-\gamma ^{2}/b_{\omega }$. The
reason why the tilting free energy cannot be truncated to $\omega ^{4}$ is
that the renormalized $b_{\omega }^{\prime }$ can become negative, in which
case a positive 6th order term is needed to stabilize the free energy. We
refer to the paper of Holakowsky for the details and only report the result
adapted to our case. The occurrence of the FE transition promotes tilting
through the biquadratic coupling term, resulting in a tilt transition at a
temperature $T$\bigskip $_{t}$ that can be also considerably higher than $%
T_{\omega }$ (which is zero in Ref. \onlinecite{Hol73}). Increasing the
magnitude of the coupling constant $\gamma $, the onset of the tilt
instability is shifted to higher temperature and the following cases are
encountered. When $b^{\prime }>$ $0$\ the tilt transition is second order
and occurs at%
\begin{equation*}
T_{t}=T_{\mathrm{C}}-\frac{a_{\omega }b}{\gamma \alpha },
\end{equation*}%
hence a temperature higher than $T_{\omega },$if $T_{\omega }\ll T_{\mathrm{C%
}}$. If $b_{0}<b^{\prime }<0$ with $b_{0}=-4\sqrt{a_{\omega }c_{\omega }/3}$
then the transition is first order and occurs between $T_{t}$ and $T_{%
\mathrm{C}}$; if $b^{\prime }<b_{0}$ the free energy may have minima at both
$P\neq 0$ and $\omega \neq 0$ (distinct from those of the pure FE phase with
$\omega =0$) already at $T_{\mathrm{C}}$ and therefore a first order
transition to a combined tilt/polar transition is possible at $T>T_{\mathrm{C%
}}$.

The above mechanism only requires that the tilt-polarization coupling is
cooperative and enough strong, which is not forbidden by any constraint of
symmetry or general principle, and we think that it can be at the basis of
the anomalous rise and merging in temperature of the tilt and FE
instabilities in Zr-rich PZT. The situation should actually be more
complicated, since the triggered transition does not produce a phase with a
clear symmetry of the polar and tilt OPs, apparently because the energy
shifts from chemical disorder compete with the energies involved in the
regular Landau expansion of a homogeneous crystal. This disorder would be
responsible for preventing the complete tilt transition with long range
order down to $T_{\mathrm{T}}\ll T_{\mathrm{IT}}$, and should somehow be
included into in the Landau expansion. These considerations should prevent
from applying the above simple formulas for deducing the magnitude of $%
\gamma $ from the upward shift of $T_{\mathrm{IT}}$. Yet, an independent
indication that $\gamma $ is indeed large and positive comes from the
positive step in the dielectric susceptibility below $T_{\mathrm{T}}$, as
shown in Sect \ref{coop}.

The possibility should be explored that a similar mechanism accounts for the
proximity, instead of crossing, of the $T_{\mathrm{MPB}}$ and $T_{\mathrm{T}%
} $ lines at $x>0.50$, as we observe in our large grain ceramic samples. In
this case, the FE OP might be the transversal component $P_{t}$, responsible
for the rotation of $\mathbf{P}$ away from the tetragonal axis, and which
takes a role in the peak of $s^{\prime }$.

The mechanism of the trigger-type transition has been applied so far to very
few cases, like Bi$_{4}$Ti$_{3}$O$_{12}$\cite{Hol75} and recently proposed
to explain the sequence of phase transitions of the multiferroic BiFeO$_{3}$,%
\cite{KB09} and is therefore considered as very rare.\cite{KB09} The
possibility that a trigger-type mechanism is also responsible for the
particular features of the phase diagrams of PZT and NBT-BT suggests that it
may be not so rare.

Finally, $T_{\mathrm{MPB}}$ seems to join also $T_{\mathrm{C}}$ smoothly,
although the upper end of the $T_{\mathrm{MPB}}$ line in the phase diagrams
above is based on only one datum and largely hypothetical (but in line with
the much more marked effect in Ref. \onlinecite{FES86}). This case is
different from the previous ones, since the OP active below $T_{\mathrm{MPB}%
} $ is not independent from that active at $T_{\mathrm{C}}$, both being the
polarization, and a triggered phase transition as above would be
meaningless. Nonetheless, the manner in which $T_{\mathrm{MPB}}$ meets $T_{%
\mathrm{C}}$ deserves further investigation.

\subsection{Competition or cooperation of tilt and polar modes?\label%
{comp/coop}}

The interpretation above contrasts with the widespread opinions that the
coupling between tilt and polar modes is negligible or competitive, and
therefore it is opportune to discuss the nature of the interaction between
polar and tilt modes.

\subsubsection{Negligible tilt-polarization coupling}

In the context of the general trends of the perovskite phase diagrams at
different compositions, the coupling between tilting and deformation or
polar modes has been considered negligible in rhombohedral perovskites.\cite%
{TB94} This assumption was based on the observation that in a large number
of rhombohedral perovskites, including FE ones, the rotation angle $\omega $
of the octahedra and the polyhedral volume ratio mentioned in Sect. \ref%
{octa-inst} are related by
\begin{equation}
V_{A}/V_{B}=6\cos ^{2}\omega -1\;,  \label{VA/VB}
\end{equation}%
which is valid in the absence of octahedral distortions. In other words, the
tilt angles are independent of the distortions,\cite{TB94} which necessarily
occur in the FE phases. This fact, however, shows that the mismatch between
the bond lengths or polyhedral volumes is accommodated almost exclusively by
the tilting mechanism, and this is understandable, since the only two normal
modes producing a large change of $V_{A}/V_{B}$ are in-phase and anti-phase
rotations of the octahedra, while all the other modes yield distortions with
little change of $V_{A}/V_{B}$.\cite{WA11} Therefore, while the validity of
Eq. (\ref{VA/VB}) is a manifestation of the fact that $V_{A}/V_{B}$ depends
almost exclusively on the tilting angle $\omega $, it does not imply a lack
of coupling between tilting and the other distortion and displacement modes,
for example influencing the type of tilt correlations.

\subsubsection{ Competitive tilt-polarization coupling}

The most widely accepted view is of a competition between FE and AFD
instabilities, based on their opposite behavior under pressure.\cite%
{SSY75,KND09,FFZ09,FAH11} Indeed, in titanates and other perovskites
pressure suppresses FE and promotes tilting, and this can be simplistically
understood in terms of a steric mechanism: FE requires more space for cation
off-centering while the octahedral rotation is promoted by a compression of
the lattice that amplifies the mismatch between B-O and A-O sublattices.
These observations, however, show that the two instabilities behave in
opposite manner under pressure, but not necessarily that they compete
against each other. It should also be noted that there are several
exceptions to the "rule" that pressure favors tilting, for example
perovskites with trivalent B = Al,\cite{RZA04} Gd,\cite{ZRA04}\ and there
are even studies that assume that the general rule is just the opposite:\
pressure induces the more symmetric cubic structure.\cite{MVJ02} These
different behaviors under pressure depend on the relative compressibilities
of the AO$_{12}$ and BO$_{6}$ polyhedra, which can be rationalized in terms
of the bond valence sum concept.\cite{AZR05} It appears that in the
zirconates and titanates the BO$_{6}$ octahedra are stiffer than the AO$%
_{12} $ cuboctahedra, which is the usual case when B has larger valence than
A.

Another strong and widely accepted\cite{FAH11} indication of competitive
AFD-FE interaction comes from simulations\cite{ZV95} on SrTiO$_{3}$ showing
that the FE mode of cation displacements along $\left\langle
111\right\rangle $ and the AFD anti-phase tilt modes compete against each
other. The simulations are done on SrTiO$_{3}$, which actually is not
ferroelectric due to quantum fluctuations, but their result is clear:
neglecting quantum fluctuations, the temperature of the tilt instability is
25\% higher if the FE mode is frozen, while $T_{\mathrm{C}}$ is 20\% higher
if the AFD mode is frozen. This means that the two modes tend to cancel each
other rather than cooperate, and the microscopic mechanism is identified in
the mutual anharmonic interaction. The behavior of $T_{\mathrm{C}}$ and $T_{%
\mathrm{IT}}$ in PZT or $T_{\mathrm{d}}$ and $T_{2}$ in NBT-NT, however,
seems the opposite. It is possible that the nature of interaction between
AFD and FE modes is different when the A ion is Pb instead of Sr, since the
difference is not only of ionic size but there is also a more strongly
covalent character of the bond when Pb goes off center.\cite{CLW96,JT02} The
other difference between the SrTiO$_{3}$ and PZT case is that in the first
only the $z$ component of the $R_{4}$ mode becomes unstable, leading to a T
structure, while in PZT all three components yield the R structure. It is
therefore unlikely that the simulation of SrTiO$_{3}$ can be plainly
generalized to Pb compounds. Rather, it may be more appropriate to refer to
similar simulations\cite{KB09} on multiferroic BiFeO$_{3}$, where also Bi$%
^{3+}$ has the stereoactive lone pair like Pb$^{2+}$. Such simulations
indicate that the sequence of FE and AFD transitions is indeed dominated by
the trigger-type mechanism, and that the coupling between the two types of
modes is strong and both competitive and cooperative, due to the fact that
the order parameters are multicomponent.\cite{KB09}

\subsubsection{Cooperative tilt-polarization coupling\label{coop}}

That the tilts are coupled with polar modes is demonstrated by the fact that
below $T_{\mathrm{T}}$ there is a positive step in the real part of the
dielectric susceptibility,\cite{VLD96,127,145} as clearly shown in Fig. \ref%
{fig dielTT}, and in the polarization.\cite{WCG78} If the modes competed
against each other, then the onset of tilting should depress rather than
enhance the polarization and its derivative, and this becomes clear when
considering the biquadratic coupling term, Eq. (\ref{Fc}), the leading term
allowed by symmetry. Below $T_{\mathrm{T}}$ the equilibrium tilt angle $%
\overline{\omega }$ starts growing, and $F_{c}$ renormalizes the term of the
free energy $\propto P^{2}$ as $a^{\prime }=a$ $-\gamma \overline{\omega }%
^{2}$ and hence the dielectric stiffness $\chi ^{-1}=$ $\partial
^{2}F/\partial P^{2}\simeq $ $a^{\prime }$. Therefore, below $T_{\mathrm{T}}$
a positive step is observed in $\chi $ if $\gamma >0$ and a negative one if $%
\gamma <0$. From Fig. \ref{fig dielTT} it appears $\gamma >0$, namely the
coupling between $\omega $ and $P$ is cooperative. In principle it would be
simple to estimate the magnitude of $\gamma $ from that of the step in $\chi
^{\prime }$ at $T_{\mathrm{T}}$, the magnitude of $\overline{\omega }$ from
diffraction and $a=\alpha \left( T-T_{\mathrm{C}}\right) $ from the
Curie-Weiss peak. However, as noted in Sect. \ref{merging}, the simple free
energy (\ref{F}) does not contain the effect of disorder that depresses $T_{%
\mathrm{T}}$, and the relevance of $\gamma $ deduced in this manner would be
questionable.

Additional indications of cooperative tilt-polarization coupling are the
fact that the application of an electric field in PZT modified with Sn and
Nb enhances $T_{\mathrm{T}}$ of few degrees.\cite{YRB04} and its prediction
in BiFeO$_{3}$ from a first-principle simulation.\cite{KB09} The coupling
between FE and AFD modes has also been discussed in relation with the
appearance of new Raman\cite{WWB11,WWB12} and infra-red\cite{BNP11} modes
below $T_{\mathrm{T}}$.

We think that the fact that tilt and polar instability lines merge over
extended composition ranges instead of crossing each other are due to a
cooperative coupling between polar/antipolar and tilt modes. It is
interesting to note that a simulation of the PZT phase diagram including
tilt degrees of freedom has already been done,\cite{KBJ06} and the $T_{%
\mathrm{MPB}}$ line, though finally crosses $T_{\mathrm{T}}$, presents a
marked bend on approaching it, exactly as appears from our anelastic and
dielectric experiments.

\subsection{Transition in the shape of the $Q^{-1}$ anomaly at $T_{\mathrm{T}%
}$: a possible sign of R/M border\label{R/M}}

As already noted in Sect. \ref{kinks}, the kink in the $T_{\mathrm{T}}$ line
and the transition in the shape of the $Q^{-1}$ anomaly (Fig. \ref{fig
TT20-53}) appear at slightly different compositions, suggesting that the
latter may have a different origin from the proximity to the MPB and hence
polarization-tilt coupling.

If this were the case, the most obvious explanation for the change of the $%
Q^{-1}$ anomaly would be the postulated border separating R and M phases.%
\cite{NCS00,SLA02,KBJ06} The existence of this border is one of the yet
unsettled issues on the phase diagram of PZT, since there are various
diffraction studies, also recent and on single crystals,\cite{YZT09,PLX10}
whose Rietveld refinements strongly suggest that the R\ and M phases coexist
at least down to $x=0.4$, so excluding a definite phase border. In addition,
according to the view that the M phase is actually a nanotwinned R or T
phase,\cite{Kha10,SSK07b} this border would not exist. Therefore, a R/M\
border would be highly significant: it would imply the existence of a long
range M phase. A puzzling feature of this border would be its verticality.
In fact, a truly vertical phase boundary in the $x-T$ phase diagram would be
understandable at a specific composition that allows a phase to be formed
with commensurate cation order. This is certainly not the case of the
postulated R/M border in PZT, where no Zr/Ti ordering has ever been
observed, and anyway $x\simeq 0.47$ is too far from the closest relevant
composition $x=\frac{1}{2}$. Indeed, the R/M boundary found by first
principles based simulations is not vertical: it starts at a triple point
with $T_{\mathrm{C}}$ and $T_{\mathrm{MPB}}$ at $x_{1}=0.463$ and ends at $%
T=0$ and $x_{2}=0.476$.\cite{KBJ06} No experimental evidence exists so far
of the crossing of such a border with change of temperature, and the change
of the shape of the $Q^{-1}$ anomaly between 0.465 and 0.48 is not a
conclusive evidence of its existence, since it might be associated with a
change of the character of the transition through polarization-tilt coupling
near the MPB. Further experiments at more closely spaced compositions are
necessary to ascertain this point.

\section{Conclusions}

Anelastic and dielectric measurements are reported at compositions of the
phase diagram of PbZr$_{1-x}$Ti$_{x}$O$_{3}$ near the two morphotropic phase
boundaries (MPB) of the rhombohedral phase with the tetragonal and the
orthorhombic phases. Several new features are found in both regions, and
discussed in terms of octahedral tilting and cooperative coupling between
the tilt and polar/antipolar modes.

We confirm the recent discovery\cite{145} of a new phase transition at a
temperature $T_{\mathrm{IT}}$ that prosecutes the border $T_{\mathrm{T}}$ of
the tilt instability up to the Curie temperature $T_{\mathrm{C}}$, in the
region where $T_{\mathrm{T}}$ drops and meets the border with the
orthorhombic antiferroelectric phase. The new phase is assumed to represent
the initial stage of octahedral tilting, without long range order due to the
enhanced cation disorder near the AFE border. Lacking evidence from
diffraction for this intermediate tilt region, the rationale for the
assumption that tilting is involved is discussed in terms of mismatch
between the networks of Pb-O and (Zr/Ti)-O bonds, as usual for tilted
perovskites. In addition, it is proposed that, due to the anisotropy of the
correlation length of different types of tilts, the initial stage of tilting
in the presence of disorder involves flat clusters of octahedra rotating
about $\left\langle 100\right\rangle $ axes, so producing $\frac{1}{2}%
\left\langle 110\right\rangle $ type modulations, even when the final long
range modulation is of $\frac{1}{2}\left\langle 111\right\rangle $ type.

The $T_{\mathrm{IT}}$ tilt instability line merges with the ferroelectric $%
T_{\mathrm{C}}$ with an evident step and both temperatures appear enhanced
with respect to the extrapolations from the region where they are far from
each other. Also the $T_{\mathrm{T}}$ line presents a clear kink when it
meets the MPB and, contrary to previous experiments, $T_{\mathrm{MPB}}$ is
found to deviate and go parallel or even merge with $T_{\mathrm{T}}$,
instead of crossing it. These observations of deviations and merging of tilt
and polar instability borders are compared to a similar example in NBT-BT,
and explained in terms of strong and cooperative interaction between the
polar and the tilt modes, which causes a trigger type transition. Since the
prevalent opinions are that tilt-polarization coupling is competitive or
negligible, and the trigger-type transitions are extremely rare, the various
indications of polar-tilt coupling in PZT are reviewed and discussed.

Another feature that is considered is a rather abrupt transition in the
shape of the anomaly in the elastic losses at $T_{\mathrm{T}}$. The anomaly
is a peak or cusp for $x\leq 0.465$ and a step for $x\geq 0.48$. The
possibility is discussed that between these two compositions there is an
actual border between rhombohedral and monoclinic phases.

\begin{acknowledgments}
The authors thank Mr. C. Capiani (ISTEC) for the skillful
preparation of the samples, Mr. P.M. Latino (ISC) and A. Morbidini
(INAF) for their technical assistance in the anelastic and
dielectric experiments.
\end{acknowledgments}

\end{document}